\documentclass[11pt,twoside]{article}
\usepackage{asp2010}

\resetcounters

\begin{document}

\title{The Taiwan Extragalactic Astronomical Data Center}
\author{S\'ebastien~Foucaud $^1$, Yasuhiro~Hashimoto $^1$, Meng-Feng~Tsai $^2$, Nicolas~Kamennoff $^3$, and the TWEA-DC team
\affil{$^1$National Taiwan Normal University, Department of Earth Sciences, 88~Tingzhou~Road, Sec. 4, Wenshan district, Taipei 11677, Taiwan}
\affil{$^2$National Central University, Department of Computer Science \& Information Engineering, 300~Jhongda~Road, Jhongli City, Taoyuan County 320, Taiwan}
\affil{$^3$Advanced Computer Science Epitech Laboratory (ACSEL), Epitech, 24~rue~Pasteur, 94270 Le Kremlin-Bic\^etre France\\
\vspace{0.1cm}e-mail: {\tt foucaud@ntnu.edu.tw}}}
\begin{abstract}
Founded in 2010, the Taiwan Extragalactic Astronomical Data Center (TWEA-DC) has for goal to propose access to large amount of data for the Taiwanese and International community, focusing its efforts on Extragalactic science. In continuation with individual efforts in Taiwan over the past few years, this is the first stepping-stone towards the building of a National Virtual Observatory.  Taking advantage of our own fast indexing algorithm (BLINK), based on a octahedral meshing of the sky coupled with a very fast kd-tree and a clever parallelization amongst available resources, TWEA-DC will propose from spring 2013 a service of "on-the-fly" matching facility, between on-site and user-based catalogs.  We will also offer access to public and private raw and reducible data available to the Taiwanese community.  
Finally, we are developing high-end on-line analysis tools, such as an automated photometric redshifts and SED fitting code (APz), and an automated groups and clusters finder (APFoF).
\end{abstract}

\section{Introduction: Data Intensive Astronomy}

The recent breakthroughs in telescopes, detectors, and also computer technology allow astronomical instruments to produce large amount of images and catalogs. It is today easier to "dial-up" a part of the sky than wait many months to have access to a telescope. With the advent of inexpensive storage technologies and the availability of high-speed networks, the concept of multi-terabyte on-line databases interoperating seamlessly is no longer outlandish. More and more catalogs are now interlinked, crossing wavelengths boundaries. Furthermore the new generation of survey telescopes (Pan-STARRS, LSST, etc) will image the entire sky every few days and yield Petabytes of data.
Over the past decade the concept of the Virtual Observatory (VO) has emerged rapidly to address challenges relating to data management, analysis, distribution and interoperability. The VO is a system in which the vast astronomical archives and databases around the world, together with analysis tools and computational services, are linked together into an integrated facility. Data centers play a central role, by providing not only a good quality service to the community (data base and software suites), but also added value based on expertise (full data analysis or research environments).

\section{The Taiwan Extragalactic Data Center: missions and goals}

The Taiwanese astronomical community is now stepping into the VO era. These efforts in Taiwan are possible through the creation of the first Taiwan based Data Center dedicated to extragalactic astronomy funded by the National Taiwan Normal University in 2010. The Taiwan Extragalactic Astronomical Data Center (TWEA-DC)\footnote{ \url{http://tweadc.es.ntnu.edu}} is designed to be a fully functioning data center around which the community can work, enabling Taiwan to join the international VO community. The efforts conducted by the VO community are already in very advanced stages, and therefore we work on the base of their latest developments, and include the available applications developed by the international community over the past decades. \\

\noindent {\bf Promoting the concept and training the new generation:}\\
One of the major goals of the TWEA-DC is to prepare the next generation of astronomers, who will have to keep up pace with the changing face of modern Astronomy. Moving into the VO era will have a dramatic impact on the existing skill base of young astronomers. Also, by making a move now in this direction, Taiwan is preparing the next generation of scientists to face the technological revolution. The project had involved several graduate students since its start, who are responsible for the different crucial stage of the build up of the DC, from the implementation of Hardware, Software, Security, and Database. Several classes have been organized during the past year to train the students to the latest techniques in Computer science. Two teams of computer scientists (from the National Central University in Taiwan, and the Advanced Computer Science Epitech Laboratory in France), including both faculty and students, are contributing fully to the TWEA-DC. In fact some graduate students from Astronomy and Computer science groups are paired and develop part of the project (such as APz, see section~3).
It is also crucial to raise awareness of the problems related to big data sets amongst Astronomers, who tend to discard these issues too easily. We organized workshops in taiwan on this topic and managed to have an increasing community gathered around the project, developing the scientific and technical cases.\\

\noindent {\bf Data Center: Database and Archive hosting:}\\
Astronomy is now based on of large datasets, covering a broad wavelength range. The challenge is to aggregate the information and generate a final product that will bridge different expertise and generate an enhanced scientific output. Large amounts of data storage are required locally to enable a fast access to images and catalogs.  \\ 
{\it Database, hosted catalogs, and peer-to-peer:} our relational database grant access to some of the largest public catalogs available (such as SDSS, 2MASS, UKIDSS, CFHTLS, SWIRE, 2DF, DEEP2, VVDS). Our DC is VO-friendly, enabling access to any dataset available in the network. We also plan to propose a peer-to-peer hosting system, where users can upload their own catalog and make it available to the community.\\
{\it Archive for Taiwanese projects:} we also aim to archive raw and reduced data for the Taiwanese community, such as Taiwanese-PI CFHT data, Lulin Observatory datasets, but also privately owned data as part of Taiwan's involvement in Pan-STARRs, Subaru-SUMIRE, or Palomar-SED machine.\\

\noindent {\bf Service-based: software development:}\\
One important mission of a Data Center is to provide the community for user-friendly tools to perform high-level analysis remotely. The large size of data available prevents users to transfer them across the network. The new generation of astronomical analysis will be conducted remotely on High Performance Computing Data Centers, creating a massive paradigm shift from our current model of astronomical research \citep[the {\it Fourth Paradigm} -][]{fourthpar}.
Some basic tools such as plotting and database queries are provided by the TWEA-DC, as well as access to some of the wide amount of tools already developed by the VO community. In addition we are developing our own set of dedicated analysis tools. Beyond the matching procedure, we will provide the community with a service that will take fully advantage of our server and eventually propose in the future a fully dedicated Domain Specific Language \citep{O18_adassxxii}.

\section{The Taiwan Extragalactic Data Center: implementations}

\noindent {\bf Current status: hardware, structure and database:}\\
The current version of TWEA-DC gathers 192Tb of data storage. The Data center has been exclusively funded by the National Taiwan Normal University.  The structure of the TWEA-DC is tailored for rapid data access and is composed of two servers, a large data storage unit and a backup System. The communication speed between the servers and the data units is of 4Gb/s, while the backup system operates at a rate of 1Gb/s. Our DC also involves an heterogenous collection of computers and we are planning in integrating some GPUs in the system. The current DC Data Warehouse structure is simple: at one end, the required data are gathered from local or on-line archives using a relational database (MySQL) and some IVOA tools : DAL (Data Access Layer) and VOTable format; at the other end, the users interface and external tools are managed by a web application. The strength of our DC is actually its middleware, standing between user interface and data: its frontend is managed as a web service and it is designed to run on parallel IT systems. The software architecture is optimized for concurrent multi-user access and to take advantage of a various kind of hardware : multi computer systems (grid, cluster, cloud), GPGPUs, etc.
We expect to release access to our database and some of the associated tools for the community in spring 2013 and our full archives towards the end of 2013.\\

\noindent {\bf The core software: Billion Line INdexing in a clicK (BLINK):}\\
BLINK is designed to index very rapidly a large amount of data according to their positions on the sky, enabling a "on-the-fly" matching \citep{blink}. This functionality is also crucial to organize data and match objects from various inputs and will be used by other software.
The indexing engine of BLINK is based on the Hierarchical Triangular Mesh \citep[HTM -][]{htm}, which describes a Quad-Tree system able to locate and identify objects on a sphere. Also, BLINK is developed to be deployed on heterogeneous parallel systems, enhancing greatly the speed of indexation, through a P2P distributed system \citep[c.f.][]{tang10}. A first version of BLINK should be available in spring 2013.
To enhance performance and flexibility, BLINK will in the future combine other indexation systems \citep[HEALPix -][]{healpix}, or even indexation not based on the position, but taking advantage of the wide range of parameters available (fluxes, shapes, compactness, etc.).\\

\noindent {\bf Automated Photometric Redshifs (APz):}\\
Redshifts are essential in any study related to galaxy evolution as well as cosmology. However computing redshifts from spectroscopy is very telescope time consuming, and will not even be possible with the next generation of large sky surveys. Fortunately, such information can be obtained using multi-band photometric information, through comparisons with a training set of spectroscopic redshifts or using template fitting.
APz is using a combination of supervised machine learning algorithms, k-Nearest Neighbor (kNN) and Support Vector Machine (SVM) to correlate color information of galaxies for which redshift is unknown to galaxies with known redshifts \citep[refer to e.g.,][]{bb10}. The strength of APz is that it will built a super training set from all information available in the database, and then train the algorithm on this dataset offline. Once the algorithm trained, users will be able to get photometric redshfits "on-the-fly" for their own catalog. APz will also provide the redshift Probability Distribution function for each individual object (pdz), and will also be able to provide physical information such as stellar masses, age, dust extinction, etc., by comparing with Synthetic Population models. APz should be available by the end of 2013.\\

\noindent {\bf Automated Probabilistic Friend-of-Friend algorithm (APFoF):}\\
Properties of galaxies correlate strongly with their close environment, which can be probed by identifying directly overdensities such as galaxy groups and clusters. However poor precision on redshifts (as obtained via photometry) prevent to conduct such study. We have developed an algorithm which take advantage of the pdz to identify overdensities from photometric redshift galaxy surveys - the probabilistic Friends-of-Friends algorithm \citep[PFoF -][]{pfof}. The parameters for PFoF require training based on catalogs of known groups and clusters which could be feed in our DC, enabling a fully optimized implementation of the algorithm. We are planning in the future to implement theoretical catalogs (from N-body simulation) within the DC, which coupled with PFoF will provide additional information (such as Dark Matter Halo masses). APFoF should be available on our DC in early 2014.\\

\acknowledgements SF acknowledges the travel support by the National Science Council of Taiwan under the grant NSC101-2914-I-003-023-A1. SF and NK are grateful to the LOC of the ADASS XXII for the support of the conference fees.

\bibliography{O14}

\end{document}